\documentclass[aps,prl,twocolumn,floatfix,superscriptaddress,showpacs]{revtex4}
\usepackage{amsmath}
\usepackage{amssymb}
\usepackage{graphicx}
\usepackage{dcolumn}
\usepackage{bm}

\begin{document}

\title{Coreless Vortices in Rotating Two-Component Quantum Droplets}

\author{H. Saarikoski}
\affiliation{Mathematical Physics, Lund Institute of Technology, 
SE-22100 Lund, Sweden}
\author{A. Harju}
\affiliation{Helsinki Institute of Physics and Department of Applied
Physics, Helsinki University of Technology, Espoo, Finland}
\author{J. Christensson}
\affiliation{Mathematical Physics, Lund Institute of Technology, 
SE-22100 Lund, Sweden}
\author{S. Bargi}
\affiliation{Mathematical Physics, Lund Institute of Technology, 
SE-22100 Lund, Sweden}
\author{M. Manninen}
\affiliation{Nanoscience Center, Department of Physics, FIN-40014,
University of Jyv\"askyl\"a, Jyv\"askyl\"a, Finland}
\author{S.~M. Reimann}
\email[Electronic address:\;]{reimann@matfys.lth.se} 
\affiliation{Mathematical Physics, Lund Institute of Technology, 
SE-22100 Lund, Sweden}

\begin{abstract}
The rotation of a quantum liquid induces vortices to carry angular
momentum. When the system is composed of multiple components that are 
distinguishable from each other, 
vortex cores in one component may be filled by 
particles of the other component, and coreless vortices form.
Based on evidence from computational methods, here we show  
that the formation of coreless vortices occurs very similarly for 
repulsively interacting bosons and fermions, 
largely independent of the form of the particle interactions. 
We further address the connection to the Halperin wave functions
of non-polarized quantum Hall states.
\end{abstract} 


\maketitle

Setting a trapped Bose-Einstein condensate (BEC) rotating, 
as for example by stirring the bosonic droplet with lasers
~\cite{fetter2008}, quantum-mechanical ``twisters'' may form 
in the cloud to carry angular momentum. Such vortices consist 
of a hole in the particle density, associated with a quantized
phase change in the order parameter around the core. 
With increasing rotation the vortices enter the bosonic cloud  
from its periphery~\cite{butts1999,kavoulakis2000} one by one, 
finally leading to the well-known vortex lattice when angular momentum
is high. The first experiment on vortices in a
BEC~\cite{matthews1999} was in fact done with a two-component gas, 
as suggested by Williams and Holland~\cite{williams1999}.  
Atoms of $^{87}$Rubidium were trapped in two distinguishable
hyperfine spin states, and vortices could be created in one component 
while the other component was at rest. 
At high rotation, the vortex lattices of the two components 
are interlaced and 'coreless' vortices form with one component filling
the vortex cores of the other 
component as a consequence of their mutual repulsion~\cite{kasamatsu2005}.
This was also experimentally observed by 
Schweikhard {\it et al.}~\cite{schweikhard2004}.

While usually attributed to superfluidity, quantized vortices as they occur  
in condensates of bosonic atoms may even appear with trapped
rotating fermion
systems~\cite{saarikoski2004,tavernier2004,toreblad2004}.
Vortex patterns in few-electron quantum dots at 
strong magnetic fields appear as hole-like quasiparticles 
in much analogy to the bosonic case~\cite{toreblad2004,manninen2005}.
Moreover, there often exists a simple transformation of the fermionic
wave function into the bosonic one~\cite{borgh2008} and
a close theoretical connection between rotating bosonic systems and
quantum Hall states~\cite{viefers2008}.
Depending on the angular momentum
vortices may either nucleate to a vortex cluster,
just like in the bosonic case~\cite{butts1999,matthews1999,madison2000},
or appear as particle-vortex composites in the finite-size precursors of
polarized~\cite{saarikoski2004,tavernier2004,toreblad2004,oaknin1995,manninen2005}
and unpolarized~\cite{merons,kiina} fractional
quantum Hall states.
In experiments rotating two-component fermion droplets may be realized with 
trapped dipolar fermion gases~\cite{baranov2008}, or with
electrons in quasi two-dimensional quantum dots~\cite{reimann2002},
where the two components are distinguished by the spin quantization
axis of the electrons, and coreless vortices are expected to
form in the regime of rapid rotation.

In this Letter, we report numerical evidence for the remarkable similarity
of coreless vortices in rotating two-component quantum droplets
regardless of the particle statistics and the form of particle interactions.

We focus on rapid rotation,
and restrict our analysis to a two-dimensional harmonic trapping 
potential of oscillator frequency $\omega $.
For a droplet composed of two distinguishable species A and B with particle
number $N=N_{\rm A} + N_{\rm B}$, the Hamiltonian is 
\begin{equation}
\hat H=\sum_{i=1}^{N_{\rm A} + N_{\rm B}} \left ( {{\mathbf p}_i^2 \over 2m} +
{1\over 2} m \omega^2 {\bf r}_i^2  \right ) +  {1\over 2} 
\sum _{i\ne j =1} ^{N_{\rm A} + N_{\rm B}} V ({\bf r}_i, {\bf r}_j)-\Omega L,
\label{Hamiltonian}
\end{equation}
where ${\bf r} = (x,y)$,
$m$ is the mass (equal for both species), $\Omega$ is the frequency
of the external rotation, and $L$ is the ($z$-component) of angular
momentum. $V$ is the particle-particle interaction
potential, which is Coulombic, 
$V= e^2/(4\pi \epsilon |{\bf r}_{i}-{\bf r}_j|)$, in the case of 
electrons in a quantum dot. 
For bosonic condensates we have used both the Coulombic potential
and the contact potential $V=U\delta({\mathbf r}_i-{\mathbf r}_j)$,
where $U=4\pi\hbar^2a/m$ and $a$ is the scattering length (here assumed to 
be the same for both components), 
which is more realistic for cold atom condensates.
In semiconductor quantum dots, the Zeeman splitting can be engineered
to be zero, as for example in 
${\rm GaAs}/{\rm Al}_x{\rm Ga}_{1-x}{\rm As}$-based
systems~\cite{saliskato2001,weisbuch1977}.
In this paper we explicitly assume that the Zeeman splitting is negligible
which leads to symmetry between the two spin components.

The eigenstates of the Hamiltonian Eq. (\ref{Hamiltonian})
are calculated by numerical diagonalization 
in the lowest Landau Level (LLL)~(for details see Refs.~\cite{toreblad2004} 
and~\cite{koskinen2007}). For large systems,
however, we applied the spin-density-functional method as described 
in Ref.~\cite{saarikoski2004}.

\begin{figure}
\includegraphics[width=0.99\columnwidth]{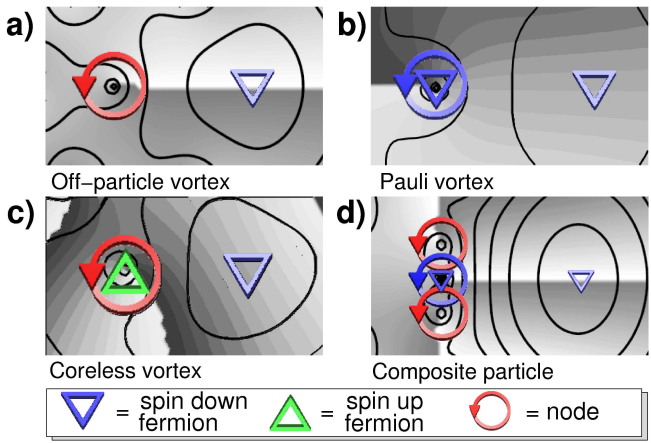}
\caption{{\it (Color online):} 
Identification of vortex structures in a restricted
wave function. The figures show RWFs for spin-$1/2$
fermions. The triangles show the fixed particle positions and 
the probing particle is marked with light blue.
Contours show the magnitude of
the RWF (on a logarithmic scale), and the gray-scale shows the phase
(darkest gray = $0$, lightest gray = $2\pi$).
a) Vortex which is
not attached to any particle. b) Pauli vortex (blue)
which is mandated by the wave function antisymmetry between interchange of
indistinguishable fermions. c) Coreless vortex where a vortex core
in one component is filled by a particle in the other component. 
d) A particle-vortex composite made up of a fermion (with a Pauli vortex) and
two additional nodes which are bound to the particle.
}
\label{sepitys}
\end{figure}

Pair-correlation functions are commonly used to analyze the structure
of the many-body wave function but integrations
average out the effect of phase singularities
which are signatures of vortices.
Therefore, we investigate the nodal structure of the wave 
function using restricted (conditional) wave functions 
(RWF)~\cite{saarikoski2004}
which essentially depict the structure of the $2N$-dimensional 
eigenstate of the Hamiltonian Eq. (\ref{Hamiltonian}), 
${\Psi(\mathbf{r}_1,\ldots,\mathbf{r}_N)}$,
in a two-dimensional subspace by fixing all but one probing particle:
\begin{equation}
\psi_\mathrm{RWF}(\mathbf{r})=
\frac{\Psi(\mathbf{r},\mathbf{r}_2^*,\dots,\mathbf{r}_N^*)}
{\Psi(\mathbf{r}_1^*,\mathbf{r}_2^*,\dots,\mathbf{r}_N^*)}.
\end{equation}
Here, the denominator is for normalization and
the fixed particle positions ${\bf r}^*_{2},\ldots, {\bf r}^*_{N} $
are determined such that the norm of $\psi_\mathrm{RWF}$ is maximized
and ${\bf r}_1^*$ is the most probable position of the probing particle.
The RWFs are two-dimensional complex functions where vortices can be 
identified directly as
nodes with a phase change of a multiple of $2\pi$ in a path around them.
Figure~\ref{sepitys} provides a guide to the identification
of vortex structures in RWFs.

In the following, we exclusively discuss equal populations of the two 
species A and B. We first study coreless vortex structures
in a 6-particle system with $N_{\rm A}=N_{\rm B}=3$, 
using the ED method in the LLL.
For a single-component fermion system the state with the
lowest angular momentum in the LLL is called the maximum-density-droplet
(MDD)~\cite{macdonald1993}, and has 
angular momentum $L_{\rm MDD}=N(N-1)/2$, which
is the minimum angular momentum compatible with the Pauli principle.
The corresponding wave function is
$\Psi_{\rm MDD}=\Pi_{i<j}^N (z_i-z_j) e^{-{1\over 2}\Sigma_k |z_k|^2}$ (with 
$z = x + iy$), where
the zeros in the polynomial are interpreted as Pauli 
vortices~\cite{saarikoski2004}. 

For single-component systems 
a simple transformation removes the attachment of Pauli vortices
to each fermion and thus leads to a symmetric wave function~\cite{xie1991}
which usually describes, to a significant degree of accuracy~\cite{borgh2008},
bosonic states at $L_{\rm Boson}=L_{\rm Fermion}-L_{\rm MDD}$.
In two-component systems Pauli vortices are only
attached to particles of the same spin and
there exist states without bosonic counterparts
at $L_{\rm Fermion}<L_{\rm MDD}$.
Thus, we consider only the angular momentum
regime $L_{\rm Fermion}\ge L_{\rm MDD}$. 

\begin{figure}
\includegraphics[width=0.99\columnwidth]{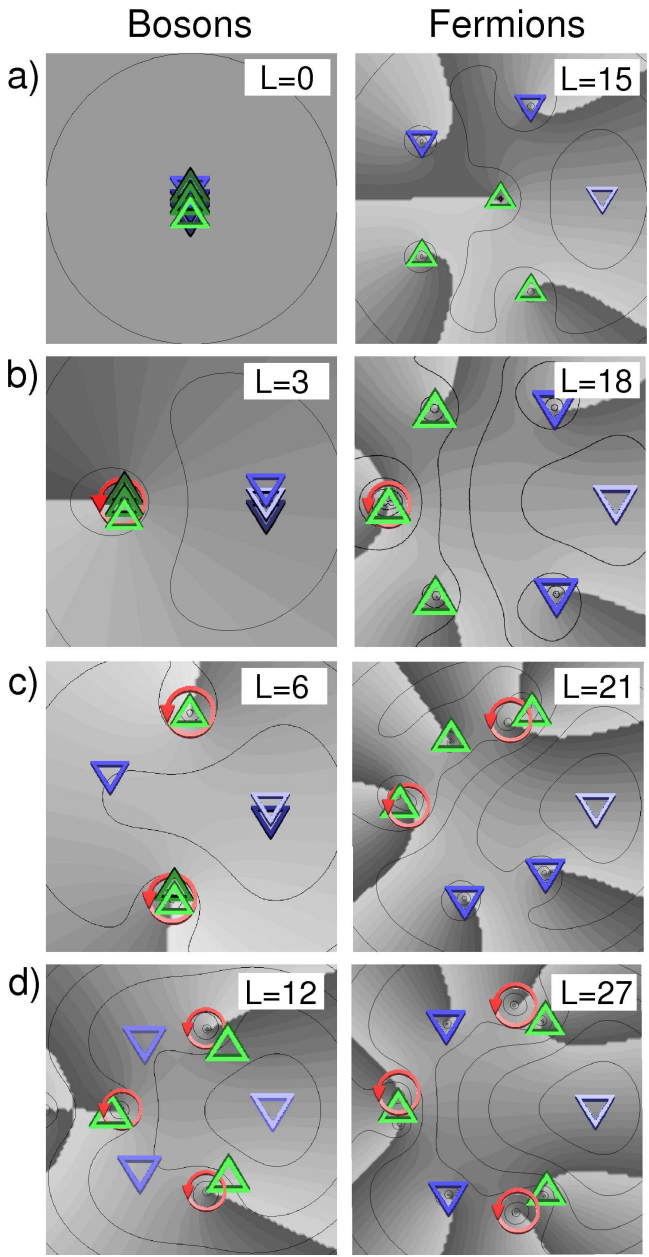}
\caption{{\it (Color online)}: The restricted wave functions
calculated with the exact diagonalization method for a two-
component system with $N_{\rm A}=N_{\rm B}=3$ for bosons at
angular momenta $L_{\rm B}$ {(\it left)} and fermionic
systems at angular momenta $L_{\rm F}=L_{\rm B}+L_{\rm MDD}=L_{\rm B}+15$
({\it right}).
The states shown are ground states in the lowest Landau level
approximation. The bosonic and fermionic states are analogous in nodal structure
when one node is removed from each fermion (these nodes are not marked
in the right panel for clarity).
The figure shows the attachment of b) one, c) two, and d) three vortices,
respectively, to particles of the opposite species
which indicates the formation of coreless vortices.
The notation is explained in the caption of Fig.~\ref{sepitys}.
}
\label{pairing}
\end{figure}

\begin{figure}
\includegraphics[width=0.99\columnwidth]{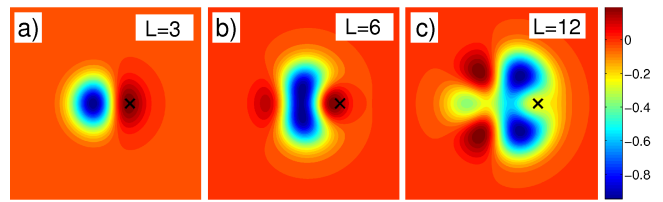}
\caption{{\it (Color online)}: Difference of pair-correlation functions 
 for two species of bosons,
for $N_{\rm A}=N_{\rm B}=3$.
a) Components are separated due to formation of the first coreless vortex.
b) Another domain forms due to formation of the second coreless vortex.
c) Formation of the third coreless vortex is associated here with
``antiferromagnetic'' ordering as seen in Fig.~\ref{pairing}d.
The reduced wave functions of corresponding states with similar
symmetries are shown in Fig.~\ref{pairing}b-d.
}
\label{paircorr}
\end{figure}

Figure~\ref{pairing} shows the RWFs of two-component bosonic
and fermionic systems for Coulomb interactions,
at ground-state angular momenta as given in each panel.  Removal of a single vortex from each fermion
position (for both components) leaves a wave function whose
nodal structure shows formation of an equal number of coreless vortices
in both fermionic and bosonic systems. These can be
identified in the RWFs as vortices
in one component that coincide with probability maxima in the opposite 
component, in order to minimize the interaction energy. 
States at intermediate angular momenta
show a vortex entry from the droplet periphery, similar to that  
in single-component systems~\cite{oaknin1995,bargi2007}.
Just as in the bosonic case, the two components of a fermion system first
separate (Fig.~\ref{pairing}b)~\cite{koskinen2007}
and, as the angular momentum increases, form interlaced vortex 
patterns (Fig.~\ref{pairing}d). 
These features in the RWFs are well reflected also 
in the pair correlations of bosonic systems (Fig.~\ref{paircorr}).
Furthermore, we find that for bosons interacting with a contact interaction
the overlaps with the corresponding states with Coulomb interaction are higher than 0.98 which shows that the coreless
vortex formation is largely independent of the form of the interaction.

There is a trend for the localization of particles to increase 
with angular momentum (and thus vorticity), as manifested by 
a narrowing of the RWF peak around the most probable particle
position (Fig.~\ref{pairing}). The effect is more clearly seen
in fermion systems where the number of vortices is higher.

\begin{figure}
\includegraphics[width=0.99\columnwidth]{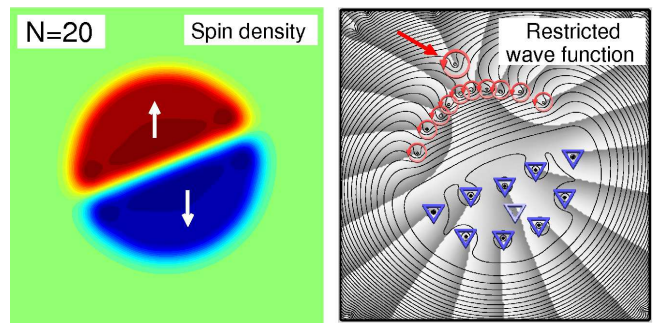}
\caption{{\it (Color online):} 
The DFT spin-density
(left) in a 20-electron droplet with equal number
of spins ($N_\uparrow=N_\downarrow=10$) shows separation of
spin components into domains. The angular momentum $L=193$ is slightly above
the MDD angular momentum $L_{\rm MDD}=190$.
The restricted wave function for the spin-down electrons (right)
shows a nodal structure with nodes equal to the number of
fixed particles and an additional coreless vortex entering the droplet (arrow).
}
\label{dftfig}
\end{figure}

For larger system sizes, the diagonalization of the 
many-body Hamiltonian Eq.~(\ref{Hamiltonian}) becomes numerically impossible.
In bosonic systems one often turns to mean-field methods
such as the Gross-Pitaevskii (GP) equation.
In the case of fermions, we can make use of density-functional theory
(DFT), as previously described for 
one-component droplets in Ref.~\cite{saarikoski2004}.    
A single-determinant wave function constructed from
the Kohn-Sham orbitals~\cite{saarikoski2005}
shows formation of coreless
vortices in much analogy to the ED results in Fig.~\ref{pairing}b.
Figure \ref{dftfig} shows the separation of the two 
components with a sharp phase boundary, analogously to the bosonic
case. For states with high angular momentum, vorticity becomes
large and the DFT solutions show interwoven vortex sheets as
previously reported
for bosonic systems using the GP equations~\cite{kasamatsu2003}.
  
\begin{figure}
\includegraphics[width=0.99\columnwidth]{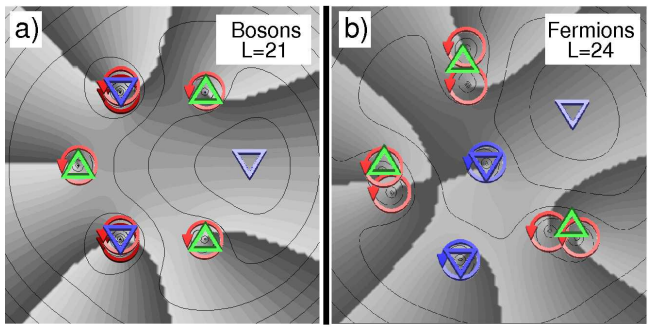}
\caption{{\it (Color online):}
a) In the regime of rapid rotation the bosonic ground state
at $L=21$ shows a mixture of both coreless vortices
and particle-double-vortex composites b)
$L=24$ fermion state with a nodal structure 
which corresponds to that of the Halperin-wave function
with $p=2$, $q=1$, which attaches two vortices on top of electrons
of the opposite spin and one (Pauli) vortex on top of electrons of
the same spin. Note that in contrast to Fig.~\ref{pairing} we
here mark all nodes.}
\label{misc}
\end{figure}

In the regime where the number of vortices in a bosonic system
exceeds the number of particles, the system shows a mixture of
coreless vortices and particle-vortex composites.
Particularly, the bosonic ground state at  $L=21$ 
shows composites of particles with two vortices in the component of the 
probing particle, and coreless vortices in the other component, as
seen in Fig.~\ref{misc}a.

The structure of the many-body states analyzed here bear much
resemblance with the trial wave functions of non-polarized quantum Hall states.
Quantum Hall states are usually polarized in high magnetic fields
due to the Zeeman-coupling but some quantum Hall plateaus, such as
$\nu={2 \over 3}$ and $\nu={2 \over 5}$, have been proposed to
be non-polarized states~\cite{chakraborty1984,xie1989}. 
The Halperin wave function for such states reads~\cite{halperin1983}
\begin{equation}
\psi =
\Pi_{i<j}^{N/2} (z_i-z_j)^q
\Pi_{k<l}^{N/2} (\tilde z_k-\tilde z_l)^q
\Pi_{m,n}^{N/2} (z_m-\tilde z_n)^p,
\label{halperinmod}
\end{equation}
where $q$ is an odd integer (due to fermion antisymmetry)
and $p$ is a positive integer
and the Gaussians have been omitted. This wave function
attaches different number of vortices on top of spin up and
spin down electrons. A vortex attached to the opposite spin component can be 
interpreted as a coreless vortex. We find that this nodal
structure has a finite-size analogue in two-component fermion droplets.
In Ref.~\cite{koskinen2007} the $L=24$ state at $N=6$ was interpreted as a
finite-size counterpart of the $\nu={2 \over 3}$ quantum Hall state
with a Halperin wave function  $p=2$, $q=1$.
Figure \ref{misc}b shows the RWF of this state where
one (Pauli) vortex is attached to each particle of the same
spin and two (coreless) vortices are attached to particles of the
opposite spin, in good agreement with the Halperin model.
However, the overlap of this state
with the Halperin wave function has been found to be low for
large particle numbers due to mixing of spin
states in the Halperin wave function (Ref. ~\cite{koskinen2007}).


We finally note that even though this study focuses on systems
with equal component sizes, coreless vortices also form with
unequal component populations. In bosonic systems these have been studied
extensively~\cite{kasamatsu2005}. 
As a fermionic example we consider results in Ref.~\cite{siljamaki2002}
where the MDD breakup was analyzed in quantum dots and 
some beyond-MDD states
were found to have partial spin polarization.
On the basis of the results presented above, these states can be
re-interpreted as vortex states in the majority spin component
whose vortex cores
are filled with an electron of the minority spin component.
Spin-selective electron transport experiments could be
able to indirectly probe for these types of coreless vortex
states in quantum dots~\cite{hitachi2006}.

This work was financially supported by the 
Swedish Research Council, the Swedish Foundation for Strategic Research, 
and NordForsk.

\end{document}